\documentstyle[12pt]{article}
\def\beq{\begin{equation}}
\def\eeq{\end{equation}}
\def\bea{\begin{eqnarray}}
\def\eea{\end{eqnarray}}
\begin{document}
\begin{titlepage}
\rightline{BROWN-HET-1169}
\rightline{hep-th/9902119}
\def\today{\ifcase\month\or
January\or February\or March\or April\or May\or June\or
July\or August\or September\or October\or November\or December\fi,
\number\year}
\vskip 1cm
\centerline{\Large \bf  Comments on}
\centerline{\Large \bf Brane Configurations with Semi-infinite D4 Branes}
\vskip 1cm
\centerline{\sc Kyungho Oh$^{a}$ and Radu Tatar$^{b}$}
\vskip 1cm
\centerline{{\it $^a$ Dept. of Mathematics, University of Missouri-St.
Louis, St. Louis, MO 63121, USA }}
\centerline{\tt oh@math.umsl.edu}
\centerline{{\it $^b$ Dept. of Physics, Brown University,
Providence, RI 02906, USA}}
\centerline{{\tt tatar@het.brown.edu}}
\vskip 2cm
\centerline{\sc Abstract}
\vskip 0.2in
We consider four dimensional supersymmetric gauge field theories from brane
configurations with the matter content given by
semi-infinite D4 branes ending on both sides of NS branes. In M theory 
configuration, we discuss the splitting of the M5 brane into 
infinite cylindrical M5 branes (which decouple) and transversal M5 brane.
The splitting condition appears naturally from the consistency of
the different projections of the Seiberg-Witten curve.

\end{titlepage}
\newpage
\section{Introduction}
\setcounter{equation}{0}
One of the main developments in  the recent years is  clarification of
 the idea that 
gauge theory and gravity are complementary descriptions
 of a single theory. 
Configurations in string/M theory have been very useful tools to study 
supersymmetric gauge field theories in different dimensions and with 
different amounts of unbroken supersymmetry. (see \cite{gk} for a detailed 
review and a complete set of references up to February 1998. For more recent 
works, see 
\cite{ura1,ura2,aot1,pa,ho,lll,ls,cp,lo,gg,ab,hu,rsu,lr,kls,klm,ka,mot}).

The usual construction of a four dimensional field theory contains D4 branes 
with one finite space direction, two NS branes (parallel for ${\cal N }= 2$ 
supersymmetry or non - parallel for ${\cal N} = 1$ supersymmetry) and possible 
D6 branes. The gauge gluons are given by strings between D4 branes. If there 
are $N$ D4 branes, the gauge group is $ SU(N)$ ( $SO(N)$ or $ Sp(N)$
 in the presence of 
an orientifold O4 or O6 plane). The matter content can be
 given by either strings 
between D4 branes and D6 branes or strings between D4 branes and semi-infinite
D4 branes ending on the NS branes \cite{wit1}. To describe the Higgs moduli
space one can use both approaches. For massive matter the discussion was 
initiated in \cite{bisky} with semi-infinite D4 branes ending on one of the NS
branes, and for massless matter the problem was solved in \cite{hoo}.

In this paper we will  discuss a brane configuration with semi-infinite
D4 branes ending on both NS branes (similar configurations are considered in
\cite{dm,dt}). The Seiberg-Witten curve is derived, and we observe that it can
be decomposed into reducible curves and this will determine a split of the
M5 brane obtained after we raise the brane configuration to M theory.

In section 2 we consider the case of the group $SU(N)$ and we explain in detail
the M5 brane splitting. In section 3 we do the same thing for the case of
orientifolds (when the gauge group is $SO(N)$ or $ Sp(N)$).

\section{$SU(N_{c})$ Gauge Theories and M5 branes Splitting}
\setcounter{equation}{0}
Consider a brane configuration for $SU(N_c)$ gauge group with
$N_f$ hypermultiplets in the fundamental representation.
In the type IIA theory on a flat space-time with time $x^0$ and space
coordinates $x^1, \ldots ,x^9$, the brane configuration consists of
two NS5 branes with worldvolume coordinates
$x^0, x^1, x^2, x^3, x^4, x^5$, $N_c$ D4 branes suspended between them
  in the $x^6$-direction,  $N_{r}$
semi-infinite D4 branes on the right of the right NS brane (which we call
 the {\em right} semi-infinite
D4 branes), and $N_{l}$
semi-infinite D4 branes on the left of the left NS brane (which we call
the {\em left} semi-infinite
D4 branes). This configuration preserves $N = 2$ supersymmetry in four
dimensions. In the models described before in the literature (see
\cite{wit1,bisky}), for simplicity, 
the semi-infinite D4 branes end on only one of the NS branes.

We are interested in  the M theory interpretation of these
brane configurations.
As usual, we introduce the complex variables:
\bea
v =x^4 + i x^5, \,w =x^8 + ix^9, \, y= \exp(- (x^6 + ix^{10})/R) 
\eea
where $x^{10}$ is the eleventh coordinate of M theory which is compactified on
a circle of radius $R$.
Now we rotate the right NS5 brane towards $w$ direction.
The new location of the right NS5 brane becomes $u:= w - \mu v = 0$ and
the left NS5 brane is still located at $w =0$. Moreover we assume that
$M_r$ of the right semi-infinite D4 branes are massless (in the sense that
they are located at $w = 0$ ) and
$M_l$ of the left semi-infinite D4 branes are massless (in the sense that they
are located at  $u = 0$).
In order to be able to rotate the  
NS5 brane, the M-theory Seiberg-Witten curve must be rational.
Since $u$ and $w$ 
 are two rational functions on this  rational curve, 
they are related by
a linear fractional transformation. Thus, after suitable constant shifts, 
we have
\bea
\label{uw}
uw= \zeta
\eea
where
$\zeta$ is a constant.
Now we project this curve to $(y, u)$-space to obtain:
\bea
\label{u}
u^{M_l}\prod_{i=1}^{N_{l}-M_l}(u - u_i) y  - P(u) = 0,
\eea
where
\bea
 P(u) = u^{N_c} + p_1  u^{N_c -1 } + \cdots + p_{N_c}
\eea
is some polynomial of degree $N_c$ because there are $N_{c}$ finite D4 branes
between the two NS branes and the the number of finite D4 branes gives the
degree of $P$.

Similarly if we project the curve to $(y, w)$-space, we get
\bea
\label{w}
Q(w) y  - Aw^{M_r} \prod_{i=1}^{N_r - M_r}(w -  w_i) = 0
\eea
where
\bea
 Q(w) = w^{N_c} + q_1  w^{N_c -1 } + \cdots + q_{N_c},
\eea
and $A$ is a normalization constant. 

In order for the equations (\ref{uw}), (\ref{u}) and (\ref{w}) 
to hold simultaneously, it is required that
\bea
P(u)Q(\zeta/u) 
 \equiv Au^{M_l}\prod_{i=1}^{N_{l}-M_l}(u - u_i)
(\zeta/u)^{M_r} \prod_{i=1}^{N_r - M_r}(\zeta/u -  w_i)
\eea 
for all $u \in {\bf C}$.

The  general solutions for P and Q are of the form
\bea
\label{solutionP}
P(u)& =& u^{N_c + M_l -N_r} P'(u)\\
\label{solutionQ}
Q(w)& =& w^{N_c +M_r -N_l}Q'(u).
\eea
assuming $N_c >N_r $ and $N_c > N_l$.
The possible solutions for $P'$ and  $Q'$ are
\bea 
P'(u) =  \prod_{i=1}^{M}(u  - u_{\alpha_i}) 
\prod_{i=1}^{N_r -M_l -M}(u - \zeta / w_{\beta_i})\\
Q'(w) = \prod_{j\neq \alpha_i} (w - \zeta/ u_j)
\prod_{j \neq \beta_i}(w - w_{j}).
\eea
If we plug (\ref{solutionP}) into (\ref{u}),  we obtain
\bea
\label{Mu}
u^{M_l}\prod_{i=1}^{N_{l}-M_l}(u - u_i) y  -
u^{N_c + M_l -N_r}\prod_{i=1}^{M}(u  - u_{\alpha_i}) 
\prod_{i=1}^{N_r -M_l -M}(u - \zeta / w_{\beta_i})  = 0,
\\
\label{Mw}
 w^{N_c +M_r -N_l} \prod_{j\neq \alpha_i} (w - \zeta/ u_j)
\prod_{j \neq \beta_i}(w - w_{j}) y  - 
Aw^{M_r} \prod_{i=1}^{N_r - M_r}(w -  w_i) = 0.
\eea
Now we can factorize these equations  into
\bea
\label{fMu}
u^{M_l}\prod_{i=1}^{M}(u  - u_{\alpha_i})\left(\prod_{j\ne \alpha_i}(u - u_j) y  -
u^{N_c  -N_r}
\prod_{i=1}^{N_r -M_l -M}(u - \zeta / w_{\beta_i})\right)  = 0
\\
\label{fMw}
w^{M_r}\prod_{j \neq \beta_i}(w - w_{j})\left( w^{N_c -N_l} \prod_{j\neq \alpha_i} (w - \zeta/ u_j)y  - 
A \prod_{i=1}^{N_r -M_l -M}(w -  w_{\beta_i})\right) = 0.
\eea
These equations suggest that the Seiberg-Witten curve in the $(u, w,y)$-space
may  be decomposed into irreducible curves.
In fact, this is possible when $M_l = M_r$ and $M= M_l + M_r$.
In this case, we can see that the
curve  $u^{M_l}\prod_{i=1}^{M}(u  - u_{\alpha_i}) = 0$ in 
$(u, y)$-space and the curve $w^{M_r}\prod_{j \neq \beta_i}(w - w_{j}) =0$
in $(w, y)$ are the images  of a curve which lies on a threefold  $uw =\zeta$.
They describe parts of the original M5 brane without the NS
branes. This means that the NS branes plus a part of the D4 branes turn into
an M5 brane as usual (which we call a transversal M5 brane) 
and the D4 branes which are lined on the opposite sides of the NS branes 
turn into M5 branes that go through the transversal M5 brane (which we call 
a cylindrical M5 brane). 
So the M5 brane is decomposed into a  cylindrical M5 brane located at
$u = w = 0$,                                                          
$M$     cylindrical M5 branes located at $u = u_{\alpha_{i}}, w = w_{i}$
and the transversal M5 brane given by the last terms in
$(\ref{fMu})$ and $(\ref{fMw})$.
This is a  phenomenon described in \cite{dm} in the discussion of the conifold
compactification. A very similar discussion appears in \cite{dt} where 
semi-infinite D4 branes ending on both sides of NS branes are also considered.
In their paper there is another type of splitting i.e. the M5 brane splits into
a flat one (without D4 branes ending on it) and a non-trivial one.
In field theory this is obtained at the root of the baryonic Higgs branch.

The third part, i.e. the transverse M5 brane, is a brane configuration
for $SU(N_c -3M_l)$ with $(N_l -M_l -M)$
flavors on the left and $(N_r -M_r -M)$ flavors on the right.
It obvious that when there is no massless flavor (i. e. $M_l = M_r = 0$),
the first two curves do not exist and, thus there is no factorization.

\section{Orientifold O6 plane, SO (Sp) Groups and M5 Brane Splitting}
\setcounter{equation}{0}

$\bullet$ $SO(2N_c)$ Case

This part will be a generalization of the case considered in
\cite{aot1} where          we have considered the case of fundamental
flavor given by semi-infinite D4 branes and we have considered only the case
  when no flavor was massless. 
Here we go one step further and we consider both massive and massless
flavor. The discussion is the same as in the $SU$ case.

To obtain the $SO$ group, we are going to introduce and orientifold
O6 plane described by:
\begin{equation}
x y = \Lambda^{4 N_c - 4 - 2N_f} v^4.
\end{equation}

Let us  rotate the left NS5 brane toward $w$ and this determines a rotation
of 
right brane in
the mirror direction (towards $-w$)
Thus   the left  NS5 brane is located at $w_{+} = 0 $ and the right NS5
brane is located at $w_{-} = 0$
where
\bea
w_\pm = w \pm \mu v.
\eea
Let $\Sigma$ be the corresponding M-theory Seiberg-Witten curve.
On $\Sigma$, the function $w_+$ will go to infinity only at one point and
$w_+$ has only a single pole there, since there is only one NS5 brane i.e.
the right NS5 brane. Thus we can identify $\Sigma$ with the 
punctured complex $w_+$-plane possibly after resolving the singularity 
at $x=y=v=0$.
Similarly, we can argue that $w_-$ has a single pole on $\Sigma$.
 Since these two functions
are rational          on a rational curve, they are related by
a   linear fractional transformation.       After suitable constant shifts,
the functions $w_\pm$ on $\Sigma$ are  related by
\bea
\label{w_+w_-}
w_+w_- = \zeta
\eea
where
$\zeta$ is a constant.
Now we project this curve to $(y, w_+)$-space. Then we obtain
\bea
\label{w_+}
w_+^{M_f}\prod_{i=1}^{N_f-M_f}(w_+ - m_i) y  - P(w_+) = 0,
\eea
where
\bea
 P(w_+) = w_+^{2N_c} + p_1  w_+^{2N_c -1 } + \cdots + p_{2N_c}
\eea
is some polynomial of degree $2N_c$.
Similarly if we project the curve to $(y, w_-)$-space, we get
\bea
\label{w_-}
Q(w_-) y  - A w_-^{M_f}\prod_{i=1}^{N_f-M_f}(w_- -  m_i) = 0
\eea
where
\bea
 Q(w_-) = w_-^{2N_c} + q_1  w_-^{2N_c -1 } + \cdots + q_{2N_c},
\eea
and $A$ is a normalization constant.
In order for the equations (\ref{w_+w_-}), (\ref{w_+}) and (\ref{w_-}) 
to hold simultaneously, it is required that
\bea
P(w_+)Q(\zeta/w_+ ) 
 \equiv Aw_+^{M_f}(\zeta/w_+)^{M_f}\prod_{i=1}^{N_f -M_f}(w_+ - m_i)
(\zeta/ w_+ -  m_i)
\eea 
for all $w_+ \in {\bf C}$.
The  solutions  will be of the form
\bea
\label{sosolutionP}
P(w_+)& =& w_+^{2N_c -N_f+M_f} \prod_{i=1}^{M}(w_+ - m_{\alpha_i}) 
\prod_{i=1}^{N_f-M_f -M}(w_+ - \zeta / m_{\beta_i})\\
\label{sosolutionQ}
Q(w_-)& =& w_-^{2N_c -N_f +M_f}\prod_{j\neq \alpha_i} (w_- - \zeta/ m_j)
\prod_{j \neq \beta_i}(w_- - m_{j})
\eea
if $2N_c \geq N_f -M_f$.
With the choice of these $P$ and $Q$, the normalization constant $A$
 will be 
\bea
A =(-1)^{N_f-M_f -M}\frac{(\zeta)^{2N_c -M}}
 {\prod_{j\neq \alpha_i} m_j \prod_{i=1}^{N_f-M_f-M} m_{\beta_i}}.
\eea 

If we plug (\ref{sosolutionP}) into (\ref{w_+}),  we obtain
\bea
\label{soMw_+}
w_+^{M_f}\prod_{i=1}^{N_f -M_f}(w_+ - m_i) y  - w_+^{2N_c -N_f +M_f} 
\prod_{i=1}^{M}(w_+ - m_{\alpha_i}) 
\prod_{i=1}^{N_f-M_f -M}(w_+ + \zeta / m_{\beta_i}) = 0\\
\label{soMw_-}
w_-^{2N_c -N_f +M_f}\prod_{j\neq \alpha_i} (w_- - \zeta/ m_j)
\prod_{j \neq \beta_i}(w_- + m_{j})y
  -A w_-^{M_f}\prod_{i=1}^{N_f-M_f}(w_- -  m_i) = 0.
\eea
After factoring out the terms $w_+^{M_f}\prod_{i=1}^{M}(w_+ - m_{\alpha_i})$
and $w_-^{M_f}\prod_{j \neq \beta_i}(w_- - m_{j})$, we obtain
\bea
\label{facMw_+}
\prod_{j\neq \alpha_i}(w_+ - m_j) y  - w_+^{2N_c -N_f+M_f}  
\prod_{i=1}^{N_f-M_f -M}(w_+ - \zeta / m_{\beta_i}) = 0\\
\label{facMw_-}
w_-^{2N_c -N_f +M_f}\prod_{j\neq \alpha_i} (w_- - \zeta/ m_j)
y  -A\prod_{i=1}^{N_f-M_f-M} (w_- -  m_{\beta_i}) = 0.
\eea
This is nothing but
a brane configuration for $SO(2N_c -M)$ theory with $N_f -M_f -M$ flavors.
The gauge group $SO(2N_c)$ with $N_f$ flavors  is broken to the gauge group
$SO(2N_c -M)$ with $N_f -M_f -M$ flavors which agrees with QCD.
In brane geometry, the  $M$ semi-infinite D4 branes are matched together with
finite D4 branes to move in $w_+ =0$ complex plane and in its mirror
complex plane. The parts of (\ref{facMw_+}) and (\ref{facMw_-}) that
decouple are again interpreted as infinite cylindrical M5 branes that go
through the transverse M5 brane.
For the case $M = 0$ we obtain the result of \cite{aot1}, so our
results are consistent.

The terms which are factored out before arriving to the formulas
(\ref{facMw_+}) and (\ref{facMw_-}) represent      the cylindrical
M5 branes which pass through the transversal M5 brane.

The extension to $SO(2N_{c} + 1)$ is trivial. In the odd case the 
equations (\ref{w_+}) and (\ref{w_-}) become:
\bea
\label{oddw_+}
\prod_{i=1}^{N_f}(w_+ - m_i) y  - P(w_+) = 0\\
\label{oddw_-}
Q(w_-) y  - A \prod_{i=1}^{N_f}(w_- +  m_i) = 0
\eea
where
\bea
P(w_+) =w_+( w_+^{2N_c} + p_1  w_+^{2N_c -1 } + \cdots + p_{2N_c})\\
Q(w_-) =w_-( w_-^{2N_c} + q_1  w_-^{2N_c -1 } + \cdots + q_{2N_c}).
\eea
and the  solutions  will be of the form
\bea
\label{oddsolutionP}
P(w_+)& =& w_+^{2N_c+1 -N_f} \prod_{i=1}^{M}(w_+ - m_{\alpha_i}) 
\prod_{i=1}^{N_f -M}(w_+ + \zeta / m_{\beta_i})\\
\label{oddsolutionQ}
Q(w_-)& =& w_-^{2N_c+1 -N_f}\prod_{j\neq \alpha_i} (w_- - \zeta/ m_j)
\prod_{j \neq \beta_i}(w_- + m_{j})
\eea
if $2N_c \geq N_f$.
When $M=0$, we have a special solution
\bea
\label{csstP1}
P(w_+)& =& w_+^{2N_c + 1 - N_f}\prod_{i=1}^{N_f}(w_+ +\zeta / m_{i})\\
\label{csstQ1}
Q(w_-)& =& w_-^{2N_c + 1 - N_f}\prod_{i=1}^{N_f} (w_- - \zeta/ m_i)
\eea
which yields 
\bea
\label{CSSTlike3}
\left(\prod_{i=1}^{N_f} m_i\right)
y \prod_{i=1}^{N_f}\left(\frac{w_-  - m_i}{w_+ + m_i}\right) =
w_+^{2N_c+1}.
\eea
This agrees   with             \cite{aot1} 
and,  thus shows the consistency of our
results. 

The discussion for $M \ne 0$ goes the same as in the $SO(2N_c)$ case and one 
obtains       the infinite cylindrical D4 branes which go through the
transverse M5 brane.

$\bullet$ $Sp(2N_c)$ Case
 
The situation is similar to $SO$ case. Now  the O6 plane is  described by:
\begin{equation}
x y = \Lambda^{2N_c - 4 - 2N_f} v^{-4}.
\end{equation}
Again the general solution is of the form
\bea
\label{spsolutionP}
P(w_+)& =& w_+^{2N_c -N_f} \prod_{i=1}^{M}(w_+ - m_{\alpha_i}) 
\prod_{i=1}^{N_f -M}(w_+ + \zeta / m_{\beta_i})\\
\label{spsolutionQ}
Q(w_-)& =& w_-^{2N_c-N_f}\prod_{j\neq \alpha_i} (w_- - \zeta/ m_j)
\prod_{j \neq \beta_i}(w_- + m_{j})
\eea
if $2N_c \geq N_f$.
For $M=0$, we get a special solution given by
\bea
xy& =&\Lambda_{N=2}^{2N_c -4 -2N_f}16 \mu^4 (w_+ - w_-)^4 \\
w_+w_- &=& \zeta\\
\label{spCSSTlike}
\left(\prod_{i=1}^{N_f} m_i \right)
y \prod_{i=1}^{N_f}\left(\frac{w_+ - m_i}{w_- + m_i}\right) &=&
w_+^{2N_c}.
\eea
On a smooth surface 
\bea
x'y' = \Lambda_{N=2}^{2N_c -4 -2N_f}
\eea 
which  maps onto the old surface via the map
$x = x'v^{-2}, y= y' v^{-2}$,
 the special solution can be described by
\bea
x'y'& =&\Lambda_{N=2}^{2N_c -4 -2N_f}\\
w_+w_- &=& \zeta\\
\label{resspCSSTlike}
\left(\prod_{i=1}^{N_f} m_i \right)
y'\prod_{i=1}^{N_f}\left(\frac{w_+ - m_i}{w_- + m_i}\right) &=&
w_+^{2N_c+2}.
\eea
If we map this curve to the old surface, then the last equation becomes:
\bea
\label{CSSTlike4}
\left(\prod_{i=1}^{N_f} m_i \right)
y\prod_{i=1}^{N_f}\left(\frac{w_+ - m_i}{w_- + m_i}\right) &=&v^{-2}
w_+^{2N_c+2}
\eea
which is the same as (5.27) of \cite{csst} after rescaling of variables.

The discussion for $M \ne 0$ is the same as before.
\section{Conclusion}
In this paper we have considered field theories obtained on the worldvolume
of D4 branes suspended between two NS branes. The matter content is given by
semi-infinite D4 branes ending on both NS branes (as opposed to the previously
considered case when they end only on one NS brane \cite{wit1,bisky}).
The Seiberg - Witten curve can be projected to $(y, u)$ and $(y, w)$ spaces
and the requirement that both projections hold simultaneously 
suggests that the Seiberg-Witten curve may be decomposed into irreducible 
curves. This implies that in M theory, the M5 brane is split into reducible
curves, one being the transversal M5 brane and the rest being infinite
cylindrical M5 branes. We have discussed the case with or without an 
orientifold
O6 plane.
\section{Acknowledgments}
We thank N. Dorey for sending us a preliminary version of his work with
D. Tong. We thank A. Hanany for discussions.
 
\end{document}